\documentclass[conference,10pt]{IEEEtran}
\usepackage{cite}

\ifCLASSINFOpdf
   \usepackage[pdftex]{graphicx}

\else

\fi

\bstctlcite{IEEEexample:BSTcontrol}

\usepackage{cite}

\usepackage{amsmath,amssymb,amsthm}

\usepackage{mathtools}
\usepackage{dblfloatfix}
\usepackage[caption=false,font=footnotesize]{subfig}
\usepackage{mathrsfs}

\newtheorem{theorem}{Theorem}

\newtheorem{lemma}[theorem]{Lemma}

\begin{document}
\title{Generalized Fano and non-Fano networks}
\author{\IEEEauthorblockN{Niladri Das and Brijesh Kumar Rai}
\IEEEauthorblockA{Department of Electronics and Electrical Engineering\\ Indian Institute of Technology Guwahati, Guwahati, Assam, India\\
Email: \{d.niladri, bkrai\}@iitg.ernet.in}}
\maketitle

\begin{abstract}
It is known that the Fano network has a vector linear solution if and only if the characteristic of the finite field is $2$; and the non-Fano network has a vector linear solution if and only if the characteristic of the finite field is not $2$. Using these properties of Fano and non-Fano networks it has been shown that linear network coding is insufficient. In this paper we generalize the properties of Fano and non-Fano networks. Specifically, by adding more nodes and edges to the Fano network, we construct a network which has a vector linear solution for any vector dimension if and only if the characteristic of the finite field belongs to an arbitrary given set of primes $\{p_1,p_2,\ldots,p_l\}$. Similarly, by adding more nodes and edges to the non-Fano network, we construct a network which has a vector linear solution for any vector dimension if and only if the characteristic of the finite field does not belong to an arbitrary given set of primes $\{p_1,p_2,\ldots,p_l\}$.
\end{abstract}

\section{Introduction}
Network coding refers to a data transmission scheme whereby instead of treating data symbols as a commodity, the intermediate nodes forward data which are functions of incoming data symbols. By using such a transmission scheme it has been shown that the min-cut upper bound on the capacity of multicast networks can be achieved \cite{li}. Linear network coding refers to the network coding scheme where all the nodes are restricted to compute only linear functions. More precisely, in linear network coding, the data symbols generated by the sources are assumed to belong to a finite field say $\mathbb{F}_q$, and nodes transmit a $\mathbb{F}_q$-linear combination of the data symbols it receives.

It is known that the linear coding capacity of a network can be dependent on the characteristic of the finite field \cite{doug}. In \cite{doug} Dougherty \textit{et al.} constructed a network, named as the Fano network, which has a scalar linear solution over any finite field of characteristic $2$, but has no vector linear solution for any vector dimension over any finite field of odd characteristic. By adding more nodes and edges to the same network we show that for any set of prime numbers $\{p_1,p_2,\ldots,p_l\}$, the resulting network has (a scalar linear solution if the characteristic belong to the given set of primes) a vector linear solution for any vector dimension if and only if the characteristic of the finite field belong to the given set.

Along with the Fano network, Dougherty \textit{et al.} also presented the non-Fano network in \cite{dougherty} which has a scalar linear solution if the characteristic of the finite field is any prime number other than $2$, but has not vector linear solution for any vector dimension if the characteristic of the finite field is $2$. We use the same network as a sub-network to construct a network which for any given set of prime numbers $\{p_1,p_2,\ldots,p_l\}$ has a vector linear solution if and only if the characteristic of the finite field does not belong to the given set.

Both of the Fano and the non-Fano networks have been constructed from the Fano and non-Fano matroid respectively. Considering the equivalence between matroids and linear network coding presented in \cite{dougherty} and \cite{kim}, our results also show that from the Fano and non-Fano matroid itself, matroids can be constructed which are representable if and only if the characteristic of the finite field belong to some arbitrary finite or co-finite set of primes. A combination of the Fano and non-Fano network has been used in \cite{doug} to show that linear network coding is insufficient. The achievable rate region for the Fano and non-Fano networks were described in reference \cite{chris}. 

It is to be noted that in references \cite{rai1} and \cite{rai2} it has been already shown that for any set of prime numbers $\{p_1,p_2,\ldots,p_l\}$, there exists a network which has a vector linear solution for any vector dimension if and only if the characteristic of the finite field belong to the given set. The authors first proved the results for sum-networks, and then showed that for every sum-network there exists a linear solvably equivalent multiple-unicast network. Similarly, for any set of prime numbers $\{p_1,p_2,\ldots,p_l\}$, they constructed a sum-network which has a vector linear solution if and only if the characteristic of the finite field does not belong to the given set. Invoking the same linear equivalence between multiple-unicast networks and sum-networks they showed the same properties hold for multiple-unicast networks. However, to the best of our knowledge this is the first time it is shown that the Fano-network and the non-Fano network itself can be used as a sub-network to construct a network having the same properties with an added desirability that these networks are simpler as they have less number of sources, terminals, nodes and edges.

The organization of the paper is as follows. In Section~\ref{s1} we reproduce the standard definition of vector linear network coding. In Section~\ref{s2}, we present the generalized Fano network, and in the following section, Section~\ref{s3}, we present the generalized non-Fano network. We conclude the paper in Section~\ref{s4}.
\begin{figure*}
\centerline{\subfloat[The Fano network]{\includegraphics[width
=0.4\textwidth]{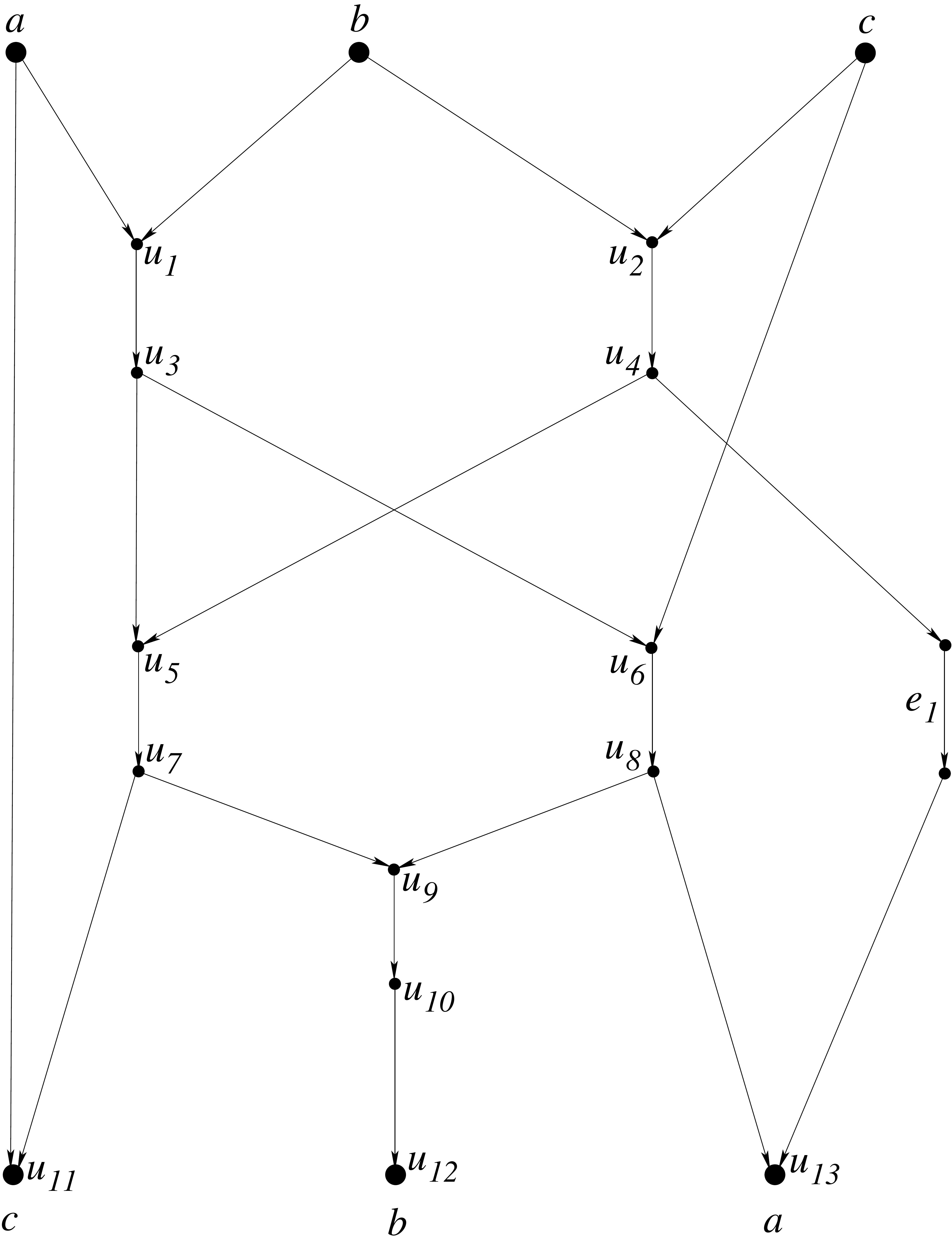}
\label{fano}}
\hfil
\subfloat[A modified Fano network]{\includegraphics[width=0.4\textwidth]{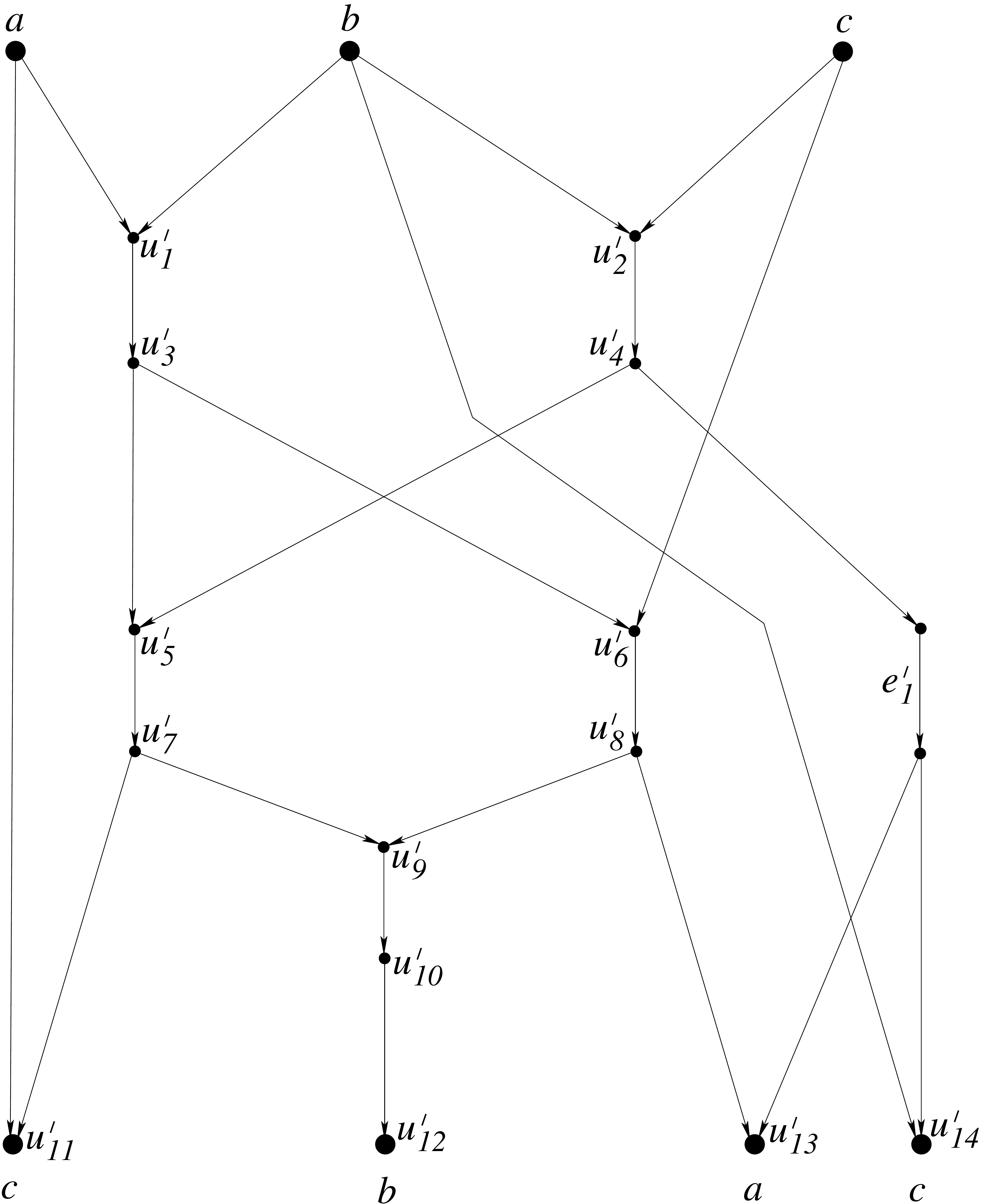}
\label{mfano}}}
\caption{A modified Fano network which has an $l$-dimensional vector linear solution if and only if the Fano network has an $l$-dimensional vector linear solution.}
\label{m}
\end{figure*}
\section{Preliminaries} \label{s1}
We represent a network by a graph $G(V,E)$. Some of the nodes are considered as sources and some as terminals. It is assumed that the sources have no incoming edges and the terminals have no outgoing edges. The set of sources is denoted by $S$; and the set of terminals is denoted by $T$. Every source generates an i.i.d. random process uniformly distributed over a finite alphabet. In linear network coding this alphabet is assumed to be a finite field $\mathbb{F}_q$. Any source process is independent of all other source processes. The source process generated by source $s\in S$ is denoted by $X_s$. Each terminal is required to retrieve the random process generated at some source. The symbol carried by an edge $e$ is denoted by $Y_e$. A node $v\in V$ is called an intermediate node if it is neither a source node nor a terminal node. $In(v)$ denotes the set of edges $e$ such that $head(e)=v$. $Out(v)$ denotes the set of edges $e$ such that $tail(e) = v$. The $i^{th}$ edge between nodes $v_1$ and $v_2$ is denoted by $(v_1,v_2,i)$. In case there is only one edge between $v_1$ and $v_2$, we denote it by $(v_1,v_2)$. 

A $k$-dimensional vector linear network coding is defined as follows. If $e\in Out(s)$ for any $s\in S$, then $Y_e = A_{s,e}X_s$ where $Y_e,X_s \in \mathbb{F}_q^k$ and $A_{s,e} \in \mathbb{F}_q^{k\times k}$. For any intermediate node $v$, if $e\in Out(v)$ and $In(v) = \{e_1,e_2,\ldots,e_n\}$, then $Y_e = A_{e_1,v}Y_{e_1} + A_{e_2,v}Y_{e_2} + \cdots + A_{e_n,v}Y_{e_n}$, where $Y_e,Y_{e_1},Y_{e_2},\ldots,Y_{e_n} \in \mathbb{F}_q^k$ and $A_{e_i,v}\in \mathbb{F}_q^{k\times k}$ for $1\leq i\leq n$. Each terminal $t\in T$ can compute a block of $k$ symbols from a source as $Y_t = A_{e_1,t}Y_{e_1} + A_{e_2,t}Y_{e_2} + \cdots + A_{e_n,t}Y_{e_n}$ where $In(t) = \{e_1,e_2,\ldots,e_n\}$, $Y_{e_1},Y_{e_2},\ldots,Y_{e_n} \in \mathbb{F}_q^k$ and $A_{e_i,t}\in \mathbb{F}_q^{k\times k}$ for $1\leq i\leq n$.

If all terminals, by obeying the above restrictions, can compute a block of $k$ symbols from their respective sources in $k$ usages of the network, then the network is said to have a $k$-dimensional vector linear solution, or analogously, the network is said to be vector linearly solvable for $k$ vector dimensions. In a $k$-dimensional vector linear solution, $k$ is called as the vector dimension. When the network has an $1$-dimensional vector linear solution the network is said to be scalar linearly solvable. The matrices indicated above by $A_{a,b}$ where $a$ and $b$ is either a node or an edge are called as local coding matrices.



\section{Generalized Fano Network}\label{s2}
In this section, for any set of prime numbers $\{p_1,p_2,\ldots,p_l\}$, a network is constructed, by adding more  nodes and edges to the Fano network, which has a vector linear solution for any message dimension if and only if the characteristic of the finite field belong to the set $\{p_1,p_2,\ldots,p_l\}$. The Fano network shown in \cite{doug} was constructed using an algorithm given in \cite{dougherty} that takes the Fano matroid as an input. Considering that there is no unique algorithm to construct a network from a matroid, and also that the construction algorithm given in \cite{dougherty} itself leaves the scope of adding more terminals, we have found that a modified Fano network (which will be shown to be linear solvably equivalent to the Fano network) to be the one which is the special case of the generalized Fano network constructed in this paper.
Towards this end, we present a modified Fano network which has an $l$-dimensional vector linear solution if and only if the Fano network has an $l$-dimensional vector linear solution. The Fano network reproduced from \cite{doug} is shown in Fig.~\ref{fano} and the modified Fano network is shown in Fig.~\ref{mfano}. Each of these two networks has three sources which generate the random processes $a,b$ and $c$ respectively. The nodes at the bottom are the terminal nodes. Each terminal demands one of the source processes from $a,b$ and $c$ as indicated in the figures. To note that the modified Fano network is constructed by adding two edges and one terminal to the Fano network.
\begin{lemma}
The network shown in Fig.~\ref{mfano} has an $l$-dimensional vector linear solution if and only if the Fano network has an $l$-dimensional vector linear solution.
\end{lemma}
\begin{IEEEproof}
We first show that if the Fano network has an $l$-dimensional vector linear solution then the modified Fano network also has an $l$-dimensional vector linear solution. Here we give the proof for the scalar linear solution, i.e., we prove that if the Fano network has a scalar linear solution then the modified Fano network also has a scalar linear solution. Say $Y_{e_1} = \alpha b + \beta c$. It we can show that $\beta \neq 0$, then upon receiving $b$ from the direct edge, node $u^\prime_{14}$ in the modified Fano network can compute $c$. Say $\beta = 0$. Since $c$ is computed at $u_{11}$, and $(u_2,u_4)$ disconnects the source that generates $c$ and node $u_{11}$, the coefficient multiplying $c$ in $Y_{(u_2,u_4)}$ cannot be zero. Hence if $\beta = 0$, it means $Y_{(u_2,u_4)}$ has been multiplied by zero, and $Y_{e_1} = 0$. Then, from $(u_6,u_8)$ solely, $a$ must be retrieved by $u_{12}$. And hence the coefficient of $b$ and $c$ in $Y_{(u_6,u_8)}$ is zero. So $b$ must be retrieved at node $u_{12}$ from $(u_5,u_7)$ solely, and this indicates that the coefficient of $c$ in $Y_{(u_5,u_7)}$ is zero. Therefore, node $u_{11}$ cannot compute $c$ which is a contradiction. So, $\beta \neq 0$. The proof for $l$-dimensional vector linear solution can be done in a similar way by taking $\beta$ as $l \times l$ matrix and showing that matrix $\beta$ has to be a full rank matrix.  

The converse that if the modified Fano network has an $l$-dimensional vector linear solution then the Fano network also has an $l$-dimensional vector linear solution is immediate because the Fano network is a sub- network of the modified Fano network.
\end{IEEEproof}
\begin{figure*}[htb]
\begin{center}
\includegraphics[width=0.95\textwidth]{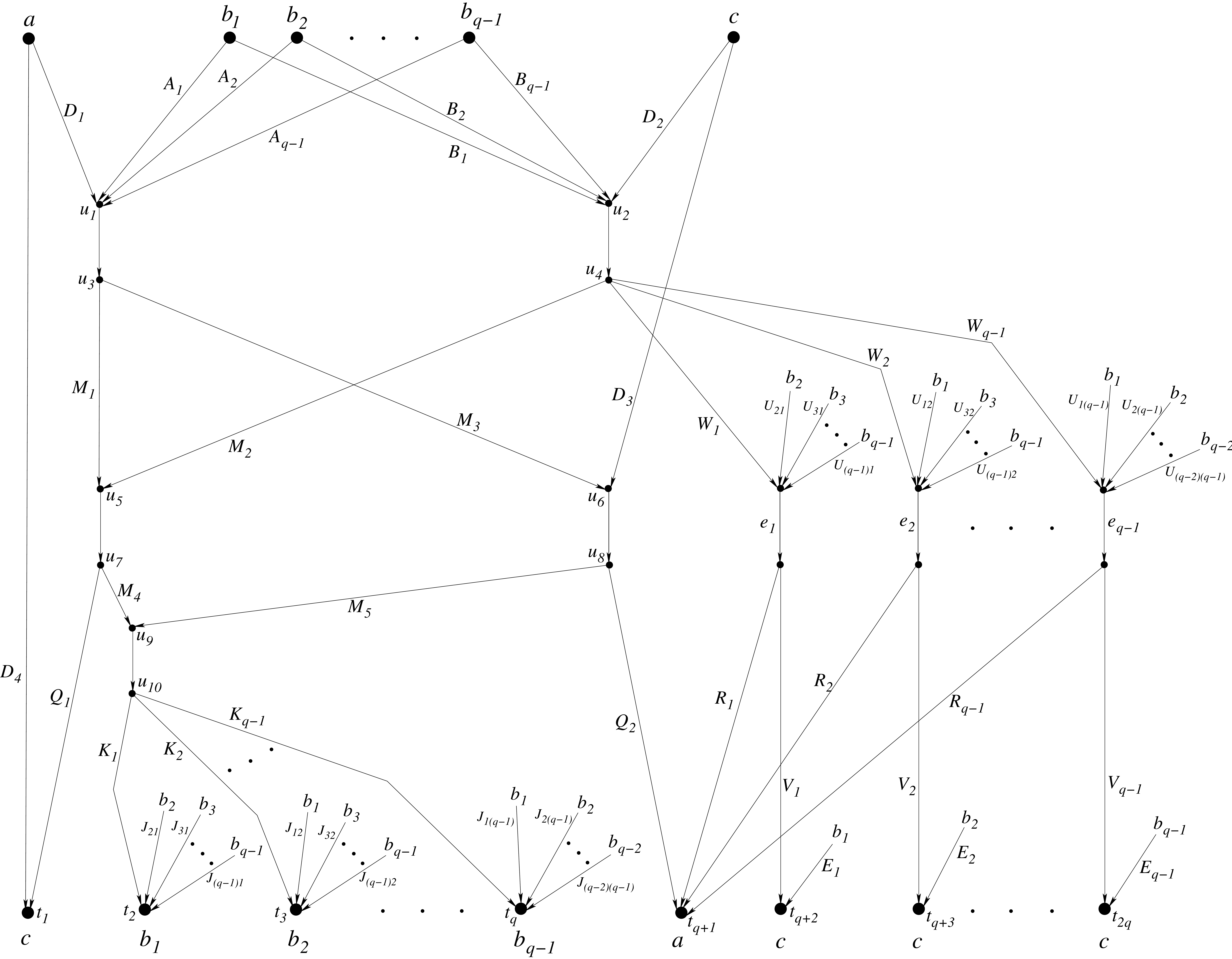}
\end{center}
\caption{Generalized Fano network: for any integer $q\geq 2$, the network is vector linearly solvable for any vector dimension if and only if the characteristic of the finite field divides $q$}
\label{genfano}
\end{figure*}
We now present the generalized Fano network in Fig.~\ref{genfano}. The top nodes are sources, and the sources generate the random processes $a,b_1,b_2,\ldots,b_{p-1}$ and $c$ respectively. The local coding matrices in the network are shown alongside the edges in capital letters. To maintain the cleanliness of the figure, some of the direct edges have been depicted by a edge without a tail node along with the notation of the source process it is connected to. The parameter $q$ which determines the number of nodes and edges in the network can take any integer value greater than or equal to two. 
Note that the network shown in Fig~\ref{genfano} reduces to the modified Fano network shown in Fig.~\ref{mfano} for $q=2$.
\begin{lemma}\label{thm1}
The network shown in Fig.~\ref{genfano} has a vector linear solution for any vector dimension if and only if the characteristic of the finite field divides $q$.
\end{lemma}
\begin{IEEEproof}
We first show that the network in Fig.~\ref{genfano} has a vector linear solution only if the characteristic of the finite field divides $q$. Say, the network has a $k$ dimensional vector linear solution.
\begin{IEEEeqnarray}{ll}
Y_{u_1,u_3}\; &= D_1a + \sum_{i=1}^{q-1}A_ib_i\\
Y_{u_2,u_4}\; &= \sum_{i=1}^{q-1}B_ib_i + D_2c\\
Y_{u_5,u_7}\; &= M_1e_{13} + M_2e_{24}\IEEEnonumber\\
&= M_1D_1a + \sum_{i=1}^{q-1} (M_1A_i + M_2B_i)b_i + M_2D_2c\\
Y_{u_6,u_8}\; &= M_3e_{13} + D_3c = M_3D_1a + \sum_{i=1}^{q-1}M_3A_ib_i + D_3c\IEEEeqnarraynumspace\\
Y_{e_i} &= W_ie_{24} + \sum_{j=1,j\neq i}^{q-1} U_{ji}b_j \quad\text{ for } 1\leq i\leq q-1\IEEEnonumber\\
&= W_iB_ib_i + \sum_{j=1,j\neq i}^{q-1} (W_iB_j + U_{ji})b_j + W_iD_2c\IEEEeqnarraynumspace\\
Y_{u_9,u_{10}}\; &= M_4e_{5,7} + M_5e_{6,8} = (M_4M_1D_1 + M_5M_3D_1)a \IEEEnonumber\\&+\> \sum_{i=1}^{q-1} (M_4(M_1A_i + M_2B_i) + M_5M_3A_i)b_i \IEEEnonumber\\&+\> (M_4M_2D_2 + M_5D_3)c
\end{IEEEeqnarray}

Since $t_1$ computes $c$, for $1\leq i\leq q-1$, we have:
\begin{IEEEeqnarray}{l}
Q_1M_1D_1 + D_4 = {0}\label{1a}\\
Q_1(M_1A_i + M_2B_i) = {0}\label{1b}\\
Q_1M_2D_2 = I\label{1c}
\end{IEEEeqnarray}

Since $t_i$ for $2\leq i\leq q$ computes $b_{i-1}$, for $1\leq i,j\leq q-1$ and $j\neq i$, we have:
\begin{IEEEeqnarray}{l}
K_i(M_4M_1D_1 + M_5M_3D_1) = {0}\label{2a}\\
K_i(M_4(M_1A_i + M_2B_i) + M_5M_3A_i) = I\label{2bi}\\
K_i(M_4(M_1A_j + M_2B_j) + M_5M_3A_j) + J_{ji} = {0}\label{2bj}\\
K_i(M_4M_2D_2 + M_5D_3) = {0}\label{2c}
\end{IEEEeqnarray}

Since $t_{q+1}$ retrieves $a$, for $1\leq i\leq q-1$, we have:
\begin{IEEEeqnarray}{l}
Q_2M_3D_1 = I\label{3a}\\
Q_2M_3A_i + R_iW_iB_i + \sum_{j=1,j\neq i}^{q-1}R_j(W_jB_i + U_{ij}) = {0}\label{3b}\IEEEeqnarraynumspace\\
Q_2D_3 + \sum_{i=1}^{q-1}R_iW_iD_2 = {0}\label{3c}
\end{IEEEeqnarray} 

Since $t_{q+1+i}$ for $1\leq i\leq q-1$ demands $c$, for $1\leq i,j\leq q-1$ and $j\neq i$, we have:
\begin{IEEEeqnarray}{l}
V_iW_iB_i + E_i = {0}\label{4bi}\\
V_i(W_iB_j + U_{ji}) = {0}\label{4bj}\\
V_iW_iD_2 = I\label{4c}
\end{IEEEeqnarray}

From equation (\ref{1c}), we know that $Q_1$ is invertible. Hence from equation (\ref{1b}), we have for $1\leq i\leq q-1$:
\begin{IEEEeqnarray}{l}
M_1A_i + M_2B_i = {0} \label{1b1}
\end{IEEEeqnarray}

From equation (\ref{2bi}), we know that $K_i$ is invertible for $1\leq i\leq q-1$. Hence from (\ref{2a}), we have,
\begin{equation}
M_4M_1D_1 + M_5M_3D_1 = {0} \label{oo1}
\end{equation}
It can be seen from equation (\ref{3a}) that $D_1$ is also invertible. Hence from equation (\ref{oo1}),
\begin{equation}
M_4M_1 + M_5M_3 = {0} \label{2a1}
\end{equation}
Substituting equation (\ref{1b1}) in equation (\ref{2bi}), we have for $1\leq i\leq q-1$:
\begin{equation}
K_iM_5M_3A_i = I\label{2bi1}
\end{equation}
Also from equation (\ref{2c}), we have:
\begin{equation}
M_4M_2D_2 + M_5D_3 = {0}\label{2c1}
\end{equation}
From equation (\ref{4c}), it can be seen that $V_i$ is invertible for $1\leq i\leq q-1$. Hence, from equation (\ref{4bj}) for $1\leq i,j\leq q-1$ and $j\neq i$,
\begin{equation}
W_iB_j + U_{ji} = {0} \label{4bj1}
\end{equation}
Substituting equation (\ref{4bj1}) in equation (\ref{3b}) for $1\leq i\leq q-1$:
\begin{equation}
Q_2M_3A_i + R_iW_iB_i = {0} \label{3b1}
\end{equation}
Since $D_2$ is invertible because of equation (\ref{1c}), from equation (\ref{3c}) we have:
\begin{equation}
Q_2D_3D_2^{-1} + \sum_{i=1}^{q-1}R_iW_i = {0} \label{3c1}
\end{equation}
Note that since from equation (\ref{2bi1}), $M_3A_i$ is invertible and from equation (\ref{3a}), $Q_2$ is invertible, $R_iW_iB_i$ in equation (\ref{3b1}) invertible, and hence $B_i$ and $A_i$ must be invertible for $1\leq i\leq q-1$. Also note that $M_2$ in equation (\ref{1c}) is invertible. Hence, from equation (\ref{1b1}), $M_1$ is invertible. Also from equation (\ref{2bi1}) $M_5$ is invertible. Substituting $R_iW_i$ from equation (\ref{3b1}) in equation (\ref{3c1}) we have:
\begin{IEEEeqnarray}{l}
Q_2D_3D_2^{-1} - \sum_{i=1}^{q-1}Q_2M_3A_iB_i^{-1} = {0} \label{3c2}\\
Q_2D_3D_2^{-1} - \sum_{i=1}^{q-1}Q_2M_3A_iB_i^{-1} = {0}\\
Q_2D_3D_2^{-1} - \sum_{i=1}^{q-1}Q_2M_3(-M_1^{-1}M_2) = {0} \quad[\text{ from } (\ref{1b1})]\IEEEeqnarraynumspace\\
Q_2D_3D_2^{-1} - \sum_{i=1}^{q-1}Q_2(-M_3M_1^{-1})M_2 = {0}\\
Q_2D_3D_2^{-1} - \sum_{i=1}^{q-1}Q_2M_5^{-1}M_4M_2 = {0}\qquad [\text{ from } (\ref{2a1})]\\
Q_2D_3D_2^{-1} - \sum_{i=1}^{q-1}Q_2(-D_3D_2^{-1}) = {0}\qquad [\text{ from } (\ref{2c1})]\\
Q_2D_3D_2^{-1} + \sum_{i=1}^{q-1}Q_2D_3D_2^{-1} = {0}\\
(q)Q_2D_3D_2^{-1} = {0}
\end{IEEEeqnarray}
Since $Q_2,D_3$ and $D_2$ are all full rank matrices, it must be that $q=0$. Now note that over any finite field of certain characteristic, $q$ is zero if and only if the characteristic divides $q$.  

We now show that the network has a scalar linear solution over a characteristic which divides $q$. Consider the following messages to be carried by the edges.
\begin{IEEEeqnarray*}{l}
Y_{u_1,u_3} = a + \sum_{i=1}^{p-1}b_i\\
Y_{u_2,u_4} = \sum_{i=1}^{q-1}b_i + c\\
Y_{u_5,u_7} = Y_{u_1,u_3} - Y_{u_2,u_4} = a - c\\
Y_{u_6,u_8} = Y_{u_1,u_3} - c = a + \sum_{i=1}^{q-1}b_i - c\\
\text{for } 1\leq i\leq q-1: \quad Y_{e_i} = b_i + c\\
Y_{u_9,u_{10}} = Y_{u_6,u_8} - Y_{u_5,u_7} = \sum_{i=1}^{q-1}b_i
\end{IEEEeqnarray*}
Now, we show that the terminals can decode their desired random variables as follows.
At terminal $t_1$, with the operation $a - Y_{u_5,u_7}$, random variable $c$ can be determined. For $1\leq i\leq q-1$, the terminal $t_{1+i}$ decodes $b_i$ as $\sum_{m=1}^{m-1}b_m - \sum_{j=1,j\neq m}^{p-1}b_j = b_i$. Since all operations are over the finite field of a characteristic which divides $q$, terminal $t_{q+1}$ decodes $a$ as $Y_{u_6,u_8} - \sum_{i=1}^{p-1}Y_{e_i} = a - qc = a$. For $1\leq i\leq q-1$, the terminal $t_{q+1+i}$  performs the operation $Y_{e_i} - b_i$ to derive $c$.
\end{IEEEproof}
\begin{center}
\begin{figure*}
\centerline{\subfloat[The non-Fano network]{\includegraphics[width
=0.4\textwidth]{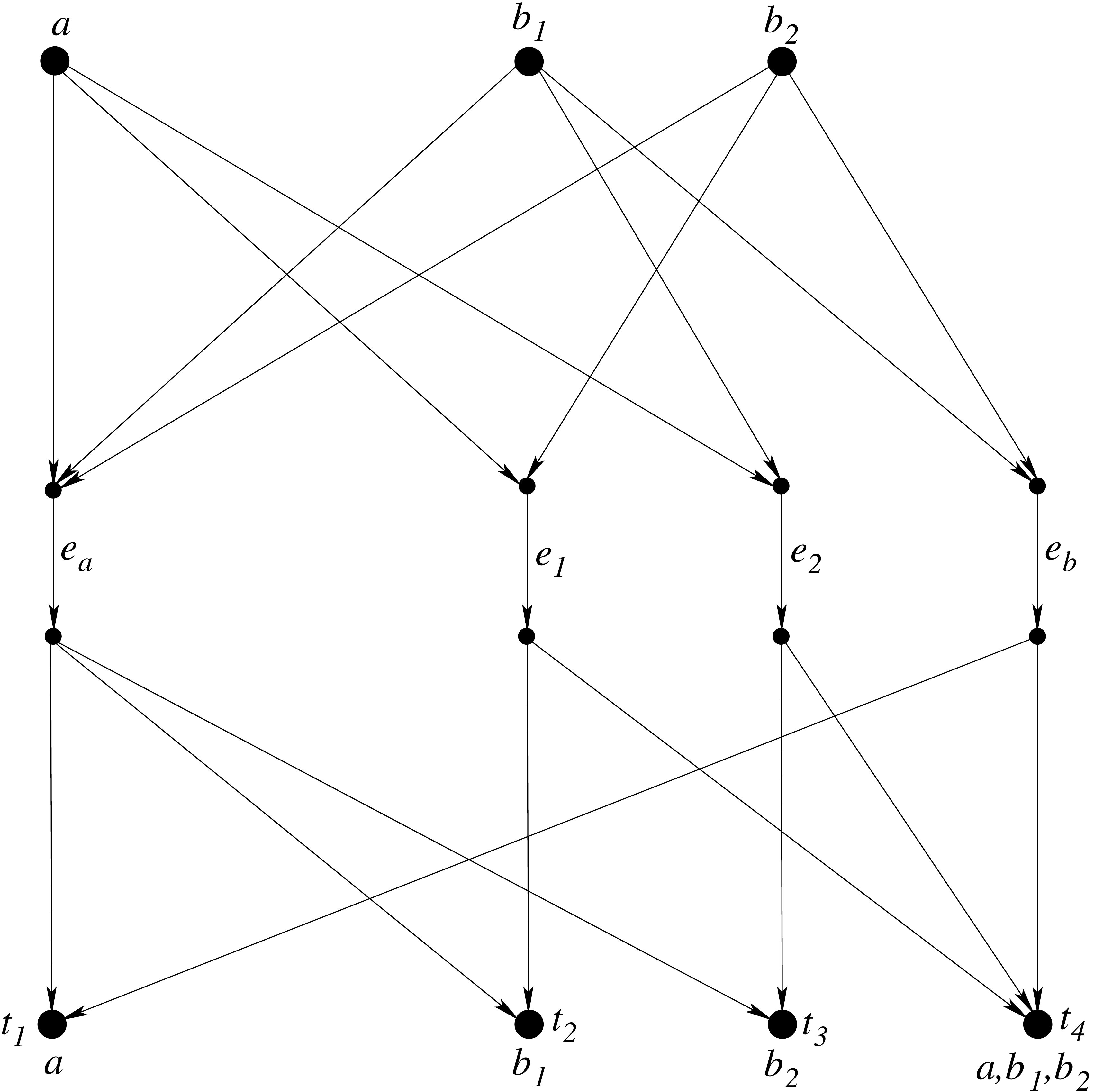}
\label{nonfano}}
\hfil
\subfloat[A modified non-Fano network]{\includegraphics[width=0.4\textwidth]{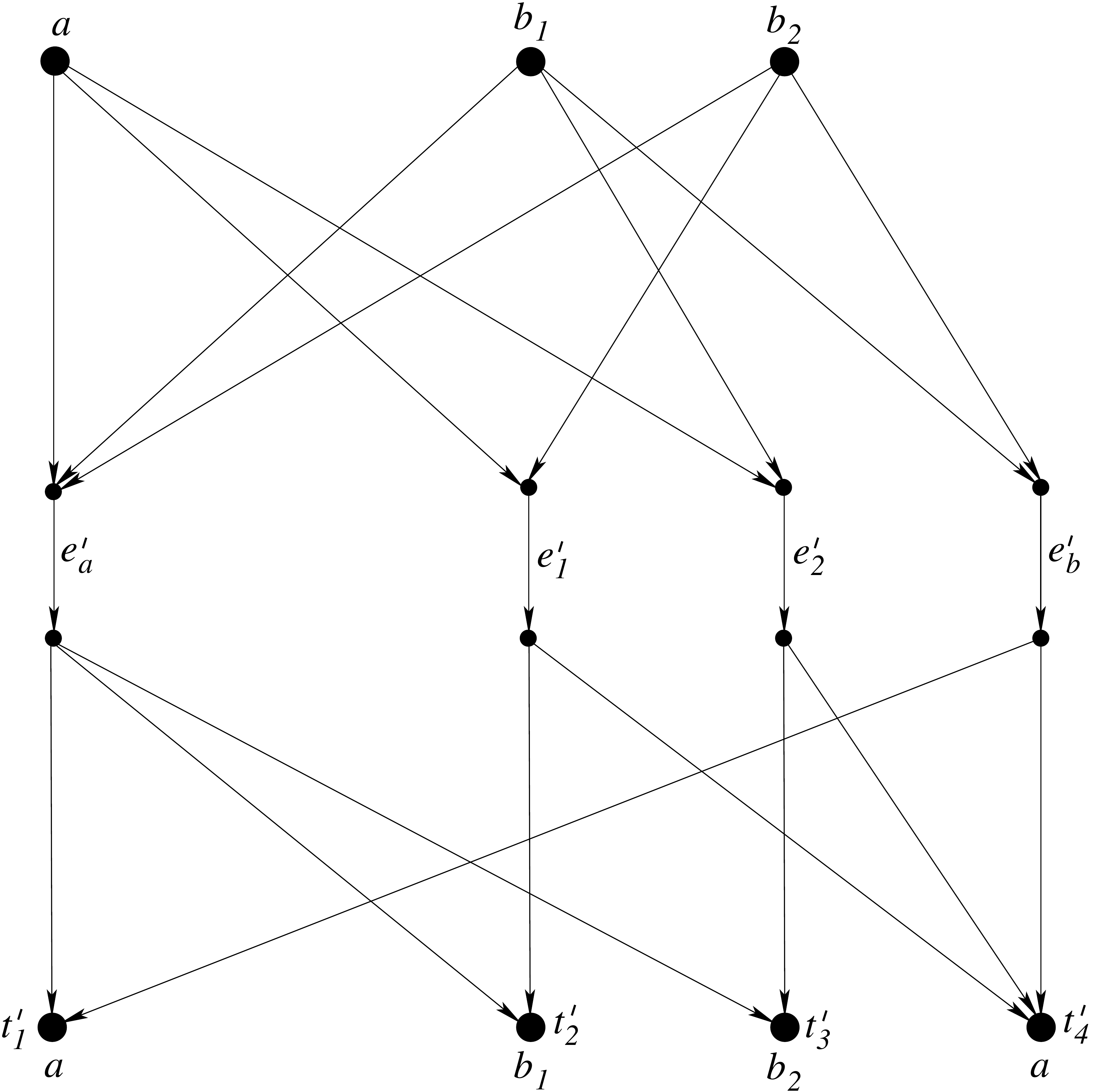}
\label{mnonfano}}}
\caption{A modified non-Fano network which has an $l$-dimensional vector linear solution if and only if the Fano network has an $l$-dimensional vector linear solution.}
\label{mnon}
\end{figure*}
\end{center}
\begin{theorem}
For any set of prime numbers $\{p_1,p_2,\ldots,p_l\}$, there exists a network, constructed by adding more nodes and edges to the Fano network, which has a vector linear solution if and only if the characteristic of the finite field belongs to the given set.
\end{theorem}
\begin{IEEEproof}
The network in Fig.~\ref{genfano} constructed for $q = p_1^{r_1}.p_2^{r_2}\ldots.p_l^{r_l}$, where $r_1,r_2,\ldots,r_l\in \mathbb{Z}^+$ is such a network.
\end{IEEEproof}

\section{Generalized non-Fano network}\label{s3}
In this section, for any set of prime numbers $\{p_1,p_2,\ldots,p_l\}$, we add more nodes and edges to the non-Fano network, to construct a network which has a vector linear solution for any vector dimension if only if the characteristic of the finite field does not belong to the given set. For this purpose, we first construct a modified non-Fano network, shown in Fig.~\ref{mnonfano}, which is linear solvably equivalent to the non-Fano network shown in \cite{dougherty} and reproduced here in Fig~\ref{nonfano}. The nodes at the top which have no incoming edges are the sources, and they not labelled to reduce clumsiness. The source process generated by a source node is indicated above the node. For any terminal, the source processes demanded by the terminal are indicated below the terminal.
\begin{lemma}
The modified non-Fano network in Fig.~\ref{mnonfano} has an $l$-dimensional vector linear solution if and only if the network in Fig.~\ref{nonfano} has an $l$-dimensional vector linear solution.
\end{lemma}
\begin{IEEEproof}
As the terminal $t^\prime_4$ in the modified non-Fano network has demands which is a subset of what $t_4$ demands in the non-Fano network, it is self-evident that if the non-Fano network has an $l$-dimensional vector linear solution then the modified non-Fano network too has an $l$-dimensional vector linear solution. We now show that if the modified non-Fano network has a scalar linear solution then the non-Fano network too has a scalar linear solution. Assume a scalar linear solution of modified non-Fano network. It can be seen that if the coefficient of $b_2$ in $Y_{e^\prime_1}$ is non-zero then $b_2$ can be retrieved from $Y_{e^\prime_1}$ by the node $t^\prime_4$ in Fig.~\ref{mnonfano} as it already knows $a$. If however, the coefficient of $b_2$ in $Y_{e^\prime_1}$ had been zero, then the coefficient of $b_2$ in $Y_{e^\prime_a}$ had also to be zero, as $Y_{e^\prime_a}$ cannot be multiplied by zero at node $t^\prime_2$ since $t^\prime_2$ needs to use the information in $Y_{e^\prime_a}$ to compute $b_1$. However, if the coefficient of $b_2$ in $Y_{e^\prime_a}$ is zero, then the node $t^\prime_3$ in Fig.~\ref{mnonfano} won't be able to compute $b_2$. Similar argument can be used to derive that the coefficient of $b_1$ in $Y_{e^\prime_2}$ is non-zero. And hence the node $t^\prime_4$ in Fig.~\ref{mnonfano} can compute all of $a,b_1$ and $b_2$.

The proof for $l$-dimensional vector linear solution can be done in a similar way.
\end{IEEEproof}
\begin{figure*}
\begin{center}
\includegraphics[width=0.68\textwidth]{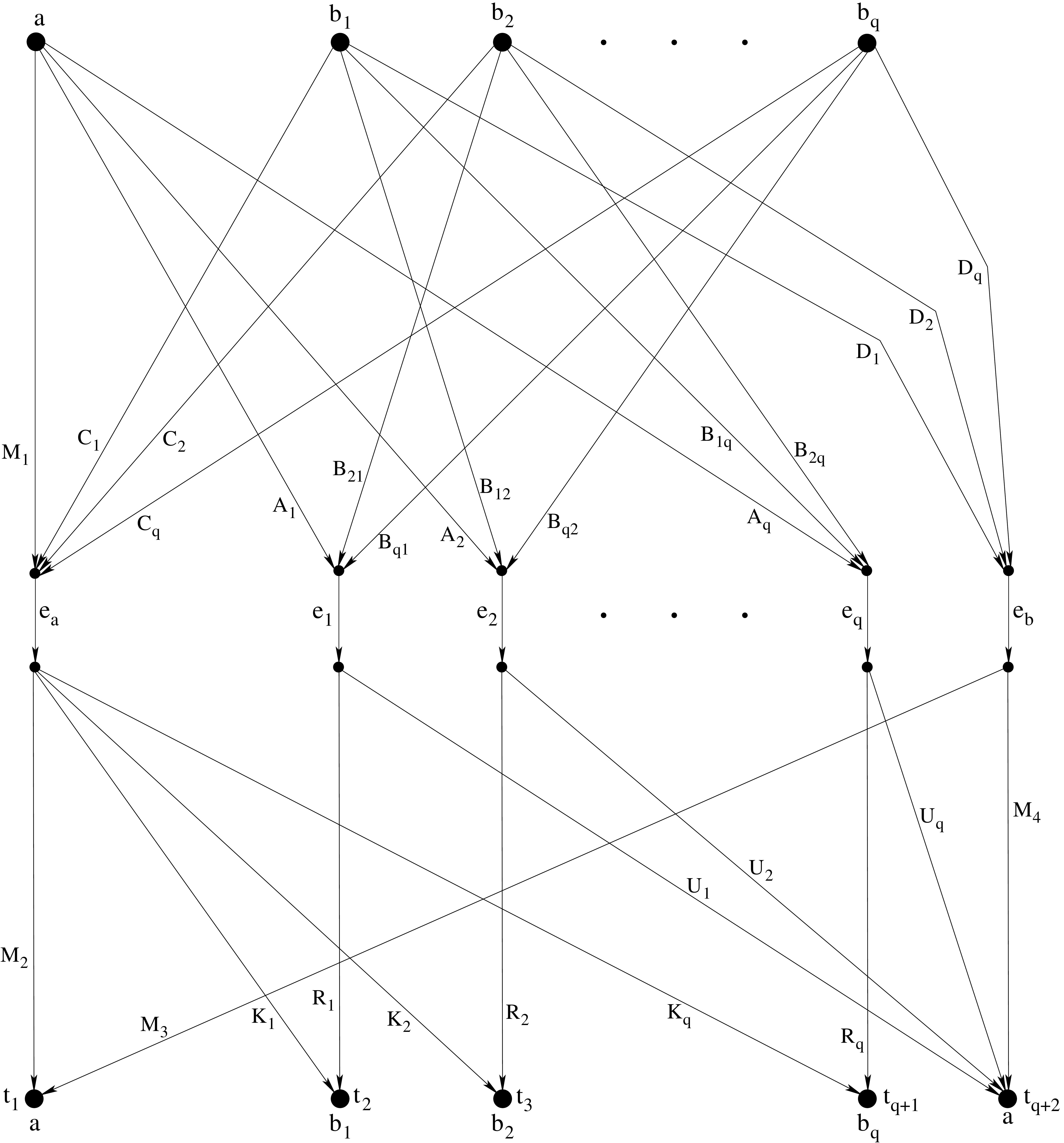}
\caption{Generalized non-Fano network: for any integer $q\geq 2$, the network is vector linearly solvable for any vector dimension if and only if the characteristic of the finite field does not divide $q$.}
\end{center}
\label{nofano}
\end{figure*}

We now present the generalized non-Fano network in Fig.~\ref{nofano}. The source nodes are not labelled in the figure for the sake of cleanliness in the diagram. $a,b_1,b_2,\ldots,b_q$ are the random processes generated by the sources. Note that for $1\leq i\leq q$, there exists no path between $tail(e_i)$ and the source that generates message $b_i$. Here also, the parameter $q$ can take any integer value greater than or equal to two. It can be verified that the network shown in Fig.~\ref{mnonfano} reduces to the modified non-Fano network shown in Fig.~\ref{mnonfano} for $q=2$.
\begin{lemma}\label{thm2}
The network shown in Fig.~\ref{nofano} has a vector linear solution for any vector dimension if and only if the characteristic of the finite field does not divide $q$.
\end{lemma}
\begin{IEEEproof}
We first list the messages carried over by the edges.
\begin{IEEEeqnarray*}{l}
Y_{e_a} = M_1a + \sum_{i=1}^q C_ib_i\\
\text{for $1\leq i \leq q$: } \quad
Y_{e_i} = A_ia + \sum_{j=1,j\neq i}^q B_{ji}b_j\\
Y_{e_b} = \sum_{i=1}^q D_ib_i\\
\end{IEEEeqnarray*}
Since, node $t_1$ computes $a$, for $1\leq i\leq q$ we have:
\begin{IEEEeqnarray}{l}
M_2M_1 = I\label{x1}\\
M_2C_i + M_3D_i = {0}\label{x2}
\end{IEEEeqnarray}
Since, node $t_{i+1}$ for $1\leq i\leq q$ computes $b_i$, for $1\leq i,j\leq q$, $j\neq i$ we have:
\begin{IEEEeqnarray}{l}
K_iM_1 + R_iA_i = {0}\label{x3}\\
K_iC_i = I\label{x4}\\
K_iC_j + R_iB_{ji} = {0}\label{x5}
\end{IEEEeqnarray}
Since, node $t_{q+2}$ computes $a$, for $1\leq i\leq q$ we have,
\begin{IEEEeqnarray}{l}
\sum_{i=1}^q U_iA_i = I\label{x6}\\
\big(\sum_{j=1,j\neq i}^q U_jB_{ij} \big)+ M_4D_i = {0}\label{x7}
\end{IEEEeqnarray}
Since, from equation (\ref{x4}), for $1\leq i\leq q$, $C_i$ is invertible, and $M_2$ is invertible from equation (\ref{x1}), $M_3D_i$ for $1\leq i\leq q$ is invertible from equation (\ref{x2}), and hence $M_3$ is invertible. Also, since both of $K_i$ and $C_i$ are invertible for $1\leq i\leq q$ because of equation (\ref{x4}), $R_i$ and $B_{ji}$ for $1\leq i,j\leq q,j\neq i$ are invertible from equation (\ref{x5}). Moreover, note that $M_1$ is invertible from equation (\ref{x1}). Now substituting equation (\ref{x2}) in equation (\ref{x7}) we get for $1\leq i\leq q$:  
\begin{IEEEeqnarray*}{l}
\big(\sum_{j=1,j\neq i}^q U_jB_{ij} \big) - M_4M_3^{-1}M_2C_i = {0}\\
-\big(\sum_{j=1,j\neq i}^q U_jR_j^{-1}K_jC_i \big) - M_4M_3^{-1}M_2C_i = {0} \quad [\text{from } \ref{x5}]\\
\big(\sum_{j=1,j\neq i}^q U_jA_jM_1^{-1}C_i \big) - M_4M_3^{-1}M_2C_i = {0}\quad [\text{from } \ref{x3}]\\
\big(\sum_{j=1,j\neq i}^q U_jA_jM_2C_i \big) - M_4M_3^{-1}M_2C_i = {0}\quad [\text{from } \ref{x1}]\\
\big(\sum_{j=1,j\neq i}^q U_jA_j - M_4M_3^{-1}\big)M_2C_i = {0}\\
\big(I - U_iA_i - M_4M_3^{-1}\big)M_2C_i = {0}\quad [\text{from } \ref{x6}]\\
I - U_iA_i - M_4M_3^{-1} = {0}\\
U_iA_i + M_4M_3^{-1} = I\\
U_iA_i = I - M_4M_3^{-1}\IEEEyesnumber\label{x8}
\end{IEEEeqnarray*}
Now, substituting equation (\ref{x8}) in equation (\ref{x6}) we get:
\begin{IEEEeqnarray*}{l}
\sum_{i=1}^q (I - M_4M_3^{-1}) = I\\
qI - qM_4M_3^{-1} = I\\
qM_4M_3^{-1} = (q-1)I \IEEEyesnumber\label{x9}
\end{IEEEeqnarray*}
Now, if the characteristic of the finite field divides $q$, then $q=0$, and equation (\ref{x9}) results into ${0} = -I$, which is a contradiction.
We now show that the network in Fig.~\ref{nofano} has a scalar linear solution if the characteristic of the finite field is does not divides $q$. Note that an element in a finite field has an inverse if and only if the characteristics of the finite field does not divide that element. Consider the following messages to be transmitted by the edges:
\begin{IEEEeqnarray*}{l}
Y_{e_a} = a + \sum_{i=1}^q b_i\\
\text{for } 1\leq i\leq q: \quad Y_{e_i} = a + \sum_{j=1,j\neq i}^q b_j\\
Y_{e_b} = \sum_{i=1}^q b_i
\end{IEEEeqnarray*}
We now show that the terminals can compute their respective demands. The terminal $t_1$ computes $a$ as $Y_{e_a} - Y_{e_b} = a$. For $1\leq i\leq q$, the terminal $t_{1+i}$ decodes $b_i$ by the operation $Y_{e_a} - Y_{e_i}$. At terminal $t_{q+2}$, since $q$ has an inverse in the finite field, $q^{-1}(\sum_{i=1}^q Y_{e_i} - (q-1)Y_{e_b}) = a$.
\end{IEEEproof}
\begin{theorem}
For any set of prime numbers $\{p_1,p_2,\ldots,p_l\}$, there exists a network, constructed by adding more nodes and edges to the non-Fano network, which has a vector linear solution if and only if the characteristic of the finite field does not belong to the given set.
\end{theorem}
\begin{IEEEproof}
The network in Fig.~\ref{nofano} constructed for $q = p_1^{r_1}.p_2^{r_2}\ldots.p_l^{r_l}$, where $r_1,r_2,\ldots,r_l\in \mathbb{Z}^+$ is such a network.
\end{IEEEproof}

\section{Conclusion}\label{s4}
The Fano and non-Fano networks have been used in the literature to show the insufficiency of linear network coding. In this paper, we have first constructed a network, named as the generalized Fano network, which for any set of primes $\{p_1,p_2,\ldots,p_l\}$, has a vector linear solution if and only if the characteristic of the finite field belongs to $\{p_1,p_2,\ldots,p_l\}$. This network reduces to the known Fano network as a special case. We have then constructed a network which for any set of primes $\{p_1,p_2,\ldots,p_l\}$, has a vector linear solution if and only if the characteristic of the finite field does not belong to $\{p_1,p_2,\ldots,p_l\}$. This network reduces to the non-Fano network as a special case.

\end{document}